\documentclass[twocolumn,a4paper]{article}

\usepackage{bm} % bold math
\usepackage{amssymb,amsfonts}
\usepackage{abstract}
\usepackage{color}
\usepackage[pdftex]{hyperref}

\begin{document}

\title{\bf{\huge{Physics of Binary Information}}}
\author{\bf{Walter Smilga} \\ Isardamm 135d, 82538 Geretsried, Germany \\ Email:\,\,wsmilga@compuserve.com }

%date{xx.xx.xxxx}

\twocolumn[
  \begin{@twocolumnfalse}
    \maketitle

\begin{abstract}
Basic concepts of theoretical particle physics, including quantum mechanics 
and Poincar\'e invariance, the leptonic mass spectrum and the proton mass, can be
derived, without reference to first principles, from intrinsic properties of the 
simplest elements of information represented by binary data.
What we comprehend as physical reality is, therefore, a reflection of 
mathematically determined logical structures, built from elements of binary data.
\end{abstract}


\vspace{1cm}

\end{@twocolumnfalse}
]

\maketitle

%{\bf PACS} 04.50.Kd, 04.60.-m, 11.30.Cp, 12.90.+b

\section{Introduction}

Throughout the last decades there has been speculation as to whether the 
basics of physics are essentially information. 
Wheeler coined the phrase ``it from bit'' \cite{jaw} and noted 
``[...]\,\,it is not unreasonable to imagine that information sits at the 
core of physics, just as it sits at the core of a computer'' \cite{jaw1}.
Von Weizs\"acker \cite{cfw1,cfw2,cfw3} and collaborators have tried to 
reconstruct physics from simple alternatives, the so-called ``ur objects'' (urs).
Related concepts are Finkelstein's space-time code \cite{df},
Penrose's spin networks \cite{rp}, and the causal spin foams of Markopoulou
and Smolin \cite{ms}.
An overview of these concepts together with additional references can be found 
in \cite{ml}.

The idea that particle physics is ultimately based on bits is supported by the 
existence of spin-1/2 particles, which can be understood as the quantum 
mechanical version of binary information, ``embedded'' in space-time.
In quantum computing, spin-1/2 particles are, in fact, used to implement 
``qubits,'' the memo\-ry elements of the quantum computers of the future.
So, at least on a ``technological'' level, a one-to-one relation between 
binary information and elementary particles has already been established.
Unfortunately, von Weiz\-s\"acker's theory of urs followed a different path 
in assuming that elementary particles ``consist of'' up to $10^{40}$ urs.
This in a way misleading assumption seems to have obstructed further 
development, so that in 1994, Lyre could state only that ``[...]\,\,investigations 
into these states and their correspondence to the known types of fundamental 
particles [...]\,\,are still underway'' \cite{hl}.

Doubts as to a correspondence with physical reality were certainly the main
reason why von Weizs\"acker's theory was largely ignored by the physics 
community.
Up to the present, this and similar concepts have been considered far 
too ``simple'' to be the basis for a realistic physical theory of the numerous
kinds of elementary particles, which in turn are subjected to up to four 
different kinds of interactions.

As far as the interactions are concerned, the situation has recently changed.
In two previous papers \cite{ws,ws1}, the author has shown that the electromagnetic 
and gra\-vi\-tational interactions can be understood as emergent properties of 
the relativistic quantum mechanics of multi-particle systems.
The reduction of these interactions to basic quantum mechanics plus Poincar\'e 
invariance means a significant reduction in the number of theoretical 
ingredients, also called ``first principles,'' in the description of physical 
reality.
It can be expected that also the remaining, strong and weak, interactions will 
find a similar explanation by Poincar\'e invariance and elementary quantum 
mechanics. 
This provokes the question as to whether the number of ingredients can be 
reduced even further, e.g., by deriving Poincar\'e invariance itself from a
more fundamental mathematical concept.

The present paper therefore attempts to understand quantum mechanics and 
Poincar\'e invariance as inherent properties of a set of binary elements 
as carriers of binary information.
This is in line with previous (futile) attempts to base physics on bits.
Now, however, we have a clear and comparatively modest goal that will guide 
our approach.
In addition, a possible path seems to be clearly marked out: by applying 
(and justifying) a quantum mechanical description to binary elements, these 
mutate into spinors, which exhibit a symmetry with respect to the rotation 
group in three dimensions SO(3).
Then we ``only'' have to give reasons for the following steps, leading from 
SO(3), via the direct product SO(3)$\times$SO(2), to the de Sitter group 
SO(3,2), and finally, by group contraction, to the Poincar\'e group P(3,1).

\section{Quantum mechanics\\and rotational symmetry}

Consider a set $B$ of $N$ binary elements $b$.
Each element $b$ may take on one of two values, e.g., ``1'' and ``0'' or ``u'' and 
``d'' (for ``up'' and ``down'').
There are, by definition, no other properties that could be used to distinguish 
one ``up'' element from another ``up'' element.
Therefore, binary elements are indistinguishable.

A very coarse property of a given set of binary elements is just the number 
of its elements.
Another property is the difference between the numbers of ``ups'' and ``downs.''

A more sophisticated way to describe a given set $B$ is to define operators 
that, applied to a vacuum state $\left| 0 \right>$, create or annihilate 
binary elements. 
The following is an excerpt from \cite{hjl}.

Let $a_k^\dagger$ and $a_k$ be the creation and annihilation operators for a binary
element in a state $k$.
These operators must satisfy the commutation rules
\begin{equation}
[ a_i, a_k^\dagger] = \delta_{ik}   \label{2-1}
\end{equation}
and
\begin{equation}
[ a_i, a_k] = [ a_i^\dagger, a_k^\dagger] = 0 . \label{2-2}
\end{equation}
A state can take on the value $u$ or $d$.

There are four possible bilinear products which do not change the number of
elements
\begin{equation}
a_u^\dagger a_d, \hspace{0.4cm} a_d^\dagger a_u, \hspace{0.4cm} 
a_u^\dagger a_u \hspace{0.3cm}\mbox{and}\hspace{0.3cm} 
a_d^\dagger a_d . \label{2-3}
\end{equation}
The first of these operators annihilates a binary element in the state $d$ and
creates a binary element in the state $u$. The second does the reverse.
The third operator counts the number of $u$ states, the fourth the number of
$d$ states.
It is convenient to divide this set of four operators into a set of three
plus the total number operator, which commutes with all the others.
\begin{eqnarray}
N &=& a_u^\dagger a_u + a_d^\dagger a_d \label{2-4}\\
\tau_+ &=& a_u^\dagger a_d \label{2-5} \\
\tau_- &=& a_d^\dagger a_u \label{2-6} \\
\tau_0 &=& \frac{1}{2} (a_u^\dagger a_u - a_d^\dagger a_d) \label{2-7}
\end{eqnarray}
The operators $\tau_+$, $\tau_-$ and $\tau_0$ satisfy 
the commutation rules
\begin{equation}
 [ \tau_0, \,\tau_+ ] \;=\;  {\tau}_+  , \\              \label{2-8}
\end{equation}
\begin{equation}
\hspace{.3cm} [ \tau_0, \,\tau_- ] \;=\; - \tau_- ,      \label{2-9}
\end{equation}
\begin{equation}
\hspace{.2cm} [ \tau_+, \tau_- ] \;=\; 2\,\tau_0 .       \label{2-10}
\end{equation}
These are the commutation relations of the special unitary group in
two dimensions, SU(2).

From $\tau_+$, $\tau_-$, and $\tau_0$, three other operators,
\begin{eqnarray}
l_1 \!&=&\! \frac{1}{2} ( \tau_+ + \tau_-), \!       \label{2-13} \\
l_2 \!&=&\! \frac{1}{2i} ( \tau_+ - \tau_-),\!       \label{2-14} \\
l_3 \!&=&\! \tau_0                                   \label{2-12}  
\end{eqnarray} 
can be constructed.
They satify the well known commutation relations of the infinitesimal 
generators of  the rotation group in three dimensions, SO(3),
\begin{equation}
[ l_1, l_2 ] = i l_3, \;\;  [l_2, l_3 ] = i l_1, \;\; [l_3, l_1 ] = i l_2.
\end{equation}
The operations of the group SO(3) are then generated by 
\begin{equation}
t = e^{i \omega_k l_k}                    \label{2-11}
\end{equation}
with three real parameters $\omega_k$.
Applying a transformation $t$ to a state $\left| u \right>$
or $\left| d \right>$ results in a linear combination 
\begin{equation}
\left| \phi \right> = c_1 \left| u \right> + c_2 \left| d \right> ,
\label{2-17}
\end{equation}
with complex coefficients $c_1$ and $c_2$.
Because the $l_k$ are Hermitian operators, $t$ is unitary.
Therefore, the {\it inner product} 
\begin{equation}
\left< \phi_1 | \phi_2 \right> \equiv c_1^* c_1 + c_2^* c_2  \label{2-18}
\end{equation}
is left invariant by these transformations, which means
\begin{equation}
c_1^* c_1 + c_2^* c_2 = 1 .
\end{equation}
Together with the inner product (\ref{2-18}), the states (\ref{2-17})
form a Hilbert space.

Considered as ``active,'' the unitary transformations (\ref{2-11}) change a 
state of the Hilbert space; considered as ``passive,'' they describe a change 
of the ``frame of reference'' without changing the informational content of 
the state.
From the ``passive'' point of view, the Hilbert space formalism makes the
description of binary states independent of the choice of a specific reference
frame. 

The real parameters $\omega_k$ form a homogeneous space, with the 
special orthogonal group SO(3) acting as a symmetry group.
The transformations (\ref{2-11}) can, therefore, be understood as 
(active or passive) {\it rotations} in a 3-dimensional parameter space.
In the Hilbert space, these rotations are represented by unitary 
transformations.
Their infinitesimal generators correspond, in the language of quantum
mechanics, to {\it observable} quantities.

Starting from a description of binary elements by creation and annihilation
operators, with the intention of making the description independent of a 
specific reference frame, we have arrived at the quantum mechanical 
description of angular momentum.
This quantum mechanical description can, therefore, be considered as a 
natural and indispensable part of any theory of binary information.

\section{Poincar\'e invariance}

In the preceding section we derived the notion of ``rotations'' of
binary elements in a 3-dimensional parameter space.
How does a time-like parameter enter into this description?

Remember that, since prehistoric times, time has been linked with the 
observation of periodic celestial events, e.g., the day--night cycle or 
the seasonal cycle.
Later on, these cycles were understood as the result of the rotation of the earth 
around its axis and of the earth around the sun, respectively.
Whereas rotations on earth can be described by parameters, belonging to 
an SO(3) symmetric parameter space, the orbit of the earth around the sun 
requires combining this parameter space with a second rotation. 
This rotation is restricted to the plane of the ecliptic and can therefore be 
described by an action of SO(2). 
Hence, a parameter space with a symmetry of SO(3)$\times$SO(2), 
describing the orientation in 3-dimensional space together with the orbital 
position, is sufficient to setup a coordinate system suitable for describing
``space and time.''
The addition of ``clocks'' with finer graduations does not change
the properties of this parameter space, as long as these clocks are 
synchronized with the ecliptical time.

An essay well worth reading about time and celestial motion has been 
written by Barbour \cite{jb}.
Our introduction of time is in line with Barbour's understanding of time
as an abstraction from the changing positions of celestial objects.

Let $x_1, x_2, x_3$ be Cartesian parameters.
Then under the operations of SO(3), the quadratic form
\begin{equation}
x_1^2 + x_2^2 + x_3^2    \label{3-1}
\end{equation}
will be constant.
Similarly, when $t_0, t_4$ are Cartesian parameters of the SO(2) part of the
parameter space, then the operations of SO(2) will leave the form
\begin{equation}
t_0^2 + t_1^2            \label{3-2}
\end{equation}
invariant.

To be able to describe coordinate transformations that connect $t$ space and 
$x$ space, it makes sense to combine both parameter sets into a five-dimensional
$t$--$x$ space.
Since the $t$  and $x$ spaces are independent parameter spaces, we are free in 
our choice of a metric connecting the $t$ and $x$ spaces.
With the metric $(++---)$, we can combine the quadratic forms (\ref{3-2}) and
(\ref{3-1}) into 
\begin{equation}
c^2 (t_0^2 + t_1^2) - (x_1^2 + x_2^2 + x_3^2) ,    \label{3-3}
\end{equation}
where $c$ is a constant with the meaning of a scaling factor. 
With this metric, special coordinate transformations, known as {\it boost}
operations, take on the simple shape of (pseudo-)rotations connecting
$x$ space and $t$ space.
The quadratic form (\ref{3-3}) is, in addition to the SO(3)$\times$SO(2) 
symmetry, invariant with respect to the full de Sitter group SO(3,2).

Note that only the subgroup SO(3)$\times$SO(2) of SO(3,2) is a symmetry
group of the ``system under consideration.''
Its extension to SO(3,2) only serves the purpose of implementing already here
certain parameter transformations that can be used to describe dynamical 
processes.
Although boost operations are not symmetry operations of the system 
under consideration, they are symmetry operations of the parameter space, 
because we have chosen a metric that renders boost operations into symmetry 
operations.
So our choice of the metric is no more than a provision intended to simplify
the mathematics.

From SO(3,2), it is only a small step to the Poincar\'e group P(3,1).
This step relies on the mathematical method of group contraction,  
formulated by In\"on\"u and Wigner \cite{iw}, who, in short, restricted 
the coordinate transformations of SO(3,2) to an infinitesimal neighborhood 
of a given point of the SO(2) orbit.
By group contraction, the SO(2) orbit is replaced by the tangent to the orbit,
which corresponds to a {\it momentum} in the direction of the orbit at the 
point of contraction.
In the contracted SO(2) system, this momentum will be denoted by $p_0$.

After contraction, the parameter space of {\it 4-momentum} has a symmetry 
group that leaves invariant the quadratic form
\begin{equation}
m^2 = p_0^2 - p_1^2 - p_2^2 - p_3^2 . \label{3-10}
\end{equation}
The symmetries of this form can be described by three rotations in 
3-dimen\-sional space, three translations in space, one translation in time, 
and three boost operations.
These operations form the {\it Poincar\'e group}, also called the {\it inhomogeneous 
Lorentz group}.
 
To summarize, Poincar\'e symmetry emerges from the attempt to provide a 
practice-oriented parametrization or ``coordinate system,'' tailored to 
the needs of an observer living on a planet of a solar system.
This does not rely on any first principle but only on usage-based decisions.
In the following, we will see that, in this parametrization, binary elements acquire 
similar properties to those on which theoretical particle physics is based.

\section{Binary elements in\\energy-momentum space}

We have seen, above, that a binary element can be described by spinor states, 
which allow a covariant description in an SO(3) symmetric parameter space. 
This concept will now be extended to obtain a covariant description of a binary 
element in the SO(3,2) symmetric parameter space.

In the following, let us represent the SU(2) algebra, defined by 
(\ref{2-8}-\ref{2-10}), by the Pauli matrices 
\begin{equation}
\sigma_{1}\!=\!\left( \begin{array}{*{2}{c}} 0 & 1 \\ 1 & 0 \end{array} \right)\!,\;
\sigma_{2}\!=\!\left( \begin{array}{*{2}{c}} 0 & -i \\ i & \;\;0 \end{array} \right)\!,\;
\sigma_{3}\!=\!\left( \begin{array}{*{2}{c}} 1 & \;\;0 \\ 0 & -1 \end{array} \right)\!.
                                                                \label{4-0}  
\end{equation}       
Then the binary elements are represented by spinors, which are the linear 
combinations of the base spinors
\begin{equation}
\left(\begin{array}{*{1}{c}} 1 \\ 0  \end{array}\right)  
\mbox{ and }
\left(\begin{array}{*{1}{c}} 0 \\ 1  \end{array}\right)  . \label{4-1}
\end{equation}

In analogy to the earth--sun example of the previous section, let us describe
the ``internal'' degrees of freedom of a binary element, as before, by an SO(3)
symmetric parameter space, and its embedding into a ``global'' coordinate system 
by another SO(3) symmetric parameter space.
Together, these parameter spaces form a space that is symmetric with respect to 
the product group SO(3)$\times$SO(3).
The restriction of the global coordinate system to SO(2) symmetric orbits and the 
additon of boost operations then result in an SO(3,2) symmetric parameter space.

The states of the product representation can then be written as linear 
combinations of four basic four-component {\it Dirac spinors} 
\begin{eqnarray}
|u_1\rangle\!&=&\!\left(\begin{array}{*{1}{c}} 1 \\ 0 \\ 0 \\ 0 \end{array}\right)  
, \;\;\;
|u_2\rangle\;\;=\;\;\left(\begin{array}{*{1}{c}} 0 \\ 1 \\ 0 \\ 0 \end{array}\right)  
                                                               \label{4-2a}\\
\nonumber \\
|v_1\rangle\!&=&\!\left(\begin{array}{*{1}{c}} 0 \\ 0 \\ 0 \\ 1 \end{array}\right)  
, \;\;\;
|v_2\rangle\;\;\,=\;\;\left(\begin{array}{*{1}{c}} 0 \\ 0 \\ 1 \\ 0 \end{array}\right)
.                                                                \label{4-2b}
\end{eqnarray}

The first group of base states (\ref{4-2a}) describes a spinor parallel to
a reference object in the SO(2) parameter space.
The second group (\ref{4-2b}) describes a spinor antiparallel to a reference 
object.
The matrix
\begin{equation}
\gamma^0 = 
\left( \begin{array}{*{2}{c}} I & \;\;0 \\ 0 & -I \end{array} \right), 
  \label{4-4}
\end{equation}
where $I$ is the $2\times 2$ unit matrix,
then delivers an eigenvalue of $+1$ if applied to the first group of spinors 
(\ref{4-2a}), and $-1$ if applied to the second (\ref{4-2b}).

This formulation is not yet covariant with respect to a transformation of
the reference frame by an action of SO(3,2).
To make it covariant, we have to find $4\times4$ matrices that transform
together with $\gamma^0$ in the same way as the reference frame.
In other words, we have to find a representation of SO(3,2) by $4\times4$
matrices.

The representation of rotations is straightforward.
Their generators are obtained from the Pauli matrices (\ref{4-0}) in the 
following form
\begin{equation}
\sigma_{ij} = \epsilon_{ijk} \left(\begin{array}{*{2}{c}} 
\sigma_k & 0 \\ 0 & \sigma_k \end{array}\right),
\; i,j,k = 1,2,3 .                                               \label{4-3} 
\end{equation}
The boost operations are generated by the $4\times4$-matrix   
\begin{equation}
\sigma^{0k} = -\sigma^{k0} = \left( \begin{array}{*{2}{c}} 0 & i\sigma_k \\
       -i\sigma_k & 0 \end{array}\right).        \label{4-6a}
\end{equation}
When we close the algebra of the matrices that we have defined so far, 
with respect to commutation products, we find the additional matrices
\begin{equation}
\gamma^k = 
\left( \begin{array}{*{2}{c}} 0 & \sigma_k \\
 -\sigma_k & 0 \end{array}\right) .                        \label{4-6}
\end{equation}
We can combine the indices $0$ and $k$ to an index $\mu = 0,\ldots,3$, and
use the metric tensor $g_{\mu\nu}=$ diag $(+1,-1,-1,-1)$ in the usual way to 
raise and lower indices.

The matrices (\ref{4-4}) and (\ref{4-6}) are Dirac's $\gamma$-matrices in the 
so-called standard or Dirac representation. 
$\gamma$-matrices satisfy the well known anti-com\-mutation relations
\begin{equation}
\{\gamma_\mu, \gamma_\nu \} = 2 g_{\mu\nu} .                  \label{4-7a}
\end{equation} 
and the commutation relations
\begin{equation}
\frac{i}{2} \, [\gamma_\mu, \gamma_\nu ] = \sigma_{\mu\nu} .  \label{4-7b}
\end{equation} 
The $4\times4$-matrices
$s_{\mu\nu}$ and $s_\mu$, built from the Dirac matrices, 
\begin{equation}
s_{\mu\nu} :=\, \frac{1}{2} \sigma_{\mu\nu}             
\mbox{   and  }  
s_\mu :=\, \frac{1}{2} \gamma_\mu                            \label{4-9}
\end{equation} 
form a representation of SO(3,2). 
The proof is by verifying the commutation relations of the 
Lie algebra of SO(3,2).

 Let us now consider a basic Dirac spinor in it's ``rest frame,'' and associate this state 
with the point $(p_0, 0, 0, 0)$ in the momentum parameter space.
Since the Dirac matrices form a representation of SO(3,2), and so also of SO(3,1), a 
Lorentz transformation of a Dirac spinor can be constructed from the appropriate Dirac matrices.
Thereby the spinor state is changed into a linear combination of all four basic spinor states,
with coefficients that are functions of the momentum parameter $p_\mu = (p_0, p_1, p_2, p_3)$, 
obtained from $(p_0, 0, 0, 0)$ by the corresponding Lorentz transformation.

Lorentz transformations leave the product $\gamma^\mu p_\mu$, applied to a Dirac spinor, 
invariant. 
The proof can be found in textbooks on relativistic quantum mechanics 
(see, e.g., \cite{sch89}).
So we end up with Dirac's equation
\begin{equation}
(\gamma_\mu p^\mu - m)\, \left| p \right> = 0 .   \label{4-10}
\end{equation}
The numerical value of $m$ is related to the value of the spin in the SO(2) 
parameter space, expressed in units of a (freely chosen) mass standard.

\section{The Pauli exclusion principle}

A simple proof of Pauli's exclusion principle, based on Bell's inequality 
\cite{jsb}, can be found in a paper by O'Hara \cite{ph}. 
The following is a (slightly modified) excerpt from this paper.

Let $\left|\psi(\lambda_1,\lambda_2)\right>$ denote a two particle state, where
$\lambda_i = (p_i, s_i)$ represents the quantum numbers of particle $i$, with
$s_i$ referring to the spin and $p_i$ to the momentum.
If the particles are in a spin-singlet state (anti-parallel spins), then their 
joint state will have the general form
\begin{equation}
\left|\psi(\lambda_1,\lambda_2)\right> = c_1 \left|u(\lambda_1)\right> \otimes\,  
                                             \left|u(\lambda_2)\right> 
                                       + c_2 \left|u(\lambda_2)\right> \otimes\, 
																			       \left|u(\lambda_1)\right> .
\end{equation}
Since the particles are in a spin-singlet state, then for the probabilities, 
the relation $P(\lambda_1\!=\!\lambda_2) \le P(s_1\!=\!s_2)\!=\!0$ holds.
Therefore, $\left<\psi(\lambda_1,\lambda_1)|\psi(\lambda_1,\lambda_1)\right>\!=\!0$ 
and hence $\left|\psi(\lambda_1,\lambda_1)\right>\!=\!0$, from the inner product
properties of a Hilbert space.
It follows that $c_1 = - c_2$, and normalizing the wave function gives 
$|c_1| =1/\sqrt{2}$.
Therefore, two binary elements, forming a joint spin-singlet state, obey 
Fermi--Dirac statistics.
A similar consideration applies if the particle spins are parallel.

\section{Multi-particle states,\\interaction,\\and entanglement}

Two Dirac spinors can be combined to a two-particle system by forming the
direct product of their single-particle Hilbert spaces.
The reduction of this product space with respect to the Poincar\'e group 
results in irreducible two-particle representations of this group.

In a previous article \cite{ws}, the author has shown that the condition of 
irreducibility leads to a correlation between two Dirac spinors, which can be 
formulated as an interaction with the properties of the electromagnetic 
interaction. 
The coupling constant calculated using that method agrees with the empirical 
value of the (low-energy) electromagnetic coupling constant.

Irreducibility is also responsible for a second interaction, which, in the 
classical limit, is described by the field equations of conformal gravity 
\cite{ws1}.
The strength of this interaction has the same order of magnitude as the strength
of the empirical gravitational force. 

Common to both interactions is that they result from the entanglement of
single-particle states, caused by the irreducibility of the two-particle representation.
This resembles the concept of spin networks, which has been studied 
by Penrose \cite{rp} and others, as a graphical method to handle the 
combination of two irreducible representations into another irreducible 
representation of the same group.

The following section illustrates the role of spin networks within the context 
of binary elements.

\section{Lepton mass relations}

About ten years ago, Gonz\'alez-Mart\'in \cite{ggm1,ggm} (G-M in the following)
obtained mass relations based on a universal structure group SL(4,R).
G-M's idea was that the structure group describes a ``substrate,'' from which
particles are generated as ``excitations'' with certain symmetric and
topological properties associated with subgroups of the structure
group.
G-M found a mass formula for the three massive leptons
\begin{equation}
m_n = 4\pi \left(\frac{16 \pi}{3}\right)^n \; m_e 
\hspace{1cm}  n = 1, 2  ,                                \label{6-8}
\end{equation}
where $m_e$ is the electron mass and $m_1$ stands for 
the muon mass, and $m_2$ for the tauon mass.
With the experimental electron mass of $0.5109989$ MeV, G-M obtained
$m_\mu = 107.5916$ MeV and $m_\tau = 1770.3$ MeV.
(The experimental values are $105.658$ and $1776.99$.)

Mathematically similar considerations to those of  G-M apply to the contraction of 
the de Sitter group SO(3,2), resulting in the ``local'' Poincar\'e group P(3,1).

Following G-M, we decompose, in a first step, a representation of SO(3,2) 
into a set of SO(3,1) representations.
Let $S$ denote the group of SO(3,2) transformations. 
Let $L$ denote the subgroup of Lorentz transformations SO(3,1) contained in 
SO(3,2), and let $P$ denote the Poincar\'e group P(3,1).

Consider a particle described by a state of the SO(3,2) symmetric Hilbert space 
$H_S$.
Assume that in a neighborhood $\cal{N}$ of the origin $\cal{O}$ this 
state is approximated by a momentum eigenstate.
When all Lorentz transformations $L$ are applied to this state,
a Hilbert space $H_L$, as a subspace of $H_S$, is obtained. 
This Hilbert space is associated with the point $\cal{O}$.

If a transformation $s \in S, s \not\in L$ is applied to a state 
of $H_L$, a new state is generated, which is not in $H_L$. 
Therefore, by applying transformations of the coset $Ls$, 
a different Hilbert space $H^s_L$ is obtained.
This Hilbert space is associated with the point $s\,\cal{O}$.
There is a one-to-one relation between cosets $Ls$ and Hilbert spaces
$H^s_L$.
The set of all cosets $Ls$ generates the total Hilbert space $H_S$.

The set of cosets forms a homogeneous space $S/L$,
where $S$ acts transitively on this space and $L$ is the isotropy group
of the origin $\cal{O}$;
the projection $\pi:\,S \rightarrow S/L$ makes $S$ a principal bundle
over $S/L$ with fiber $L$. 

Adding up all the different $H^s_L$s means an integration over the
homogeneous space $S/L$.
This integral delivers a decomposition of $H_S$ into a sum of 
different $H^s_L$,
\begin{equation}
H_S = \int d\Omega\;H^s_L = \int ds\,\frac{d\Omega}{ds}\;H^s_L , \label{6-1}
\end{equation}
where $d\Omega$ is the infinitesimal volume element in $S/L$.
The Jacobian $d\Omega/ds$ is a measure of the number of 
different Hilbert spaces $H^s_L$ obtained by an infinitesimal 
transformation $ds$.
With a properly chosen parametrization, such that $\int \! ds = 1$, the 
Jacobian becomes identical to the volume $V(S/L)$ of $S/L$.
The volume of $S/L$ has been determined in \cite{ggm1}:
\begin{equation}
V(S/L) = \frac{16 \pi}{3} .                              \label{6-7}
\end{equation}

Now consider a Dirac spinor within a spin network of binary elements.
Since the momentum of a Dirac spinor has three independent components,
we can use each of them to entangle the spinor state with the states of
up to three other objects of the spin network.
In this case, the Hilbert space $H_S$ of the spinor kinematics relative to
the other objects is the direct product of up to three Hilbert spaces, 
$H^{(1)}_S, H^{(2)}_S$, and $H^{(3)}_S$.
With the decomposition (\ref{6-1}) of each Hilbert space, the integrals then
contain products of the factor $V(S/L)$.
The following products correspond, respectively, to one, two, and three objects.
\begin{equation}
\left(\frac{16 \pi}{3}\right) \mbox{,} \;\;
\left(\frac{16 \pi}{3}\right)^2  \;\; \mbox{ and } \;\;
\left(\frac{16 \pi}{3}\right)^3                          \label{6-9}
\end{equation}

Restricting the resulting Hilbert spaces to those Hilbert spaces that are 
associated with the point $\cal{O}$, means dividing the factors in 
(\ref{6-9}) by $V(S/L)$.
This results in
\begin{equation}
1  \mbox{ , } \;\;
\left(\frac{16 \pi}{3}\right)  \;\; \mbox{ and } \;\;
\left(\frac{16 \pi}{3}\right)^2  .                      \label{6-10}
\end{equation}

Next, recall that spinors are described by representations of P(3,1) in the
neighborhood $\cal{N}$ of $\cal{O}$, rather than of SO(3,1) at $\cal{O}$.
Representations of P(3,1) in $\cal{N}$ are obtained from representations 
of SO(3,1) at $\cal{O}$ by adding infinitesimal transformations $t \in S$.
By applying all $t$ to $L$, the cosets $Lt$ are obtained. 
They form a homogeneous space $P/L$ with a volume \cite{ggm} of
\begin{equation}
V(P/L)\;=\;V(U(1))\;=\;4\pi .                          \label{6-11}
\end{equation}
This factor corresponds to an additional integration over the homogeneous 
space $P/L$.
It applies to those Hilbert spaces that are obtained by reduction of 
product representations with respect to SO(3,1).
The Hilbert space $H_L$ associated with the point $\cal{O}$ does not 
require this factor, because it is already a representation of P(3,1). 

Multiplying the terms (\ref{6-10}) by the appropriate factors results 
in 
\begin{equation}
1  \mbox{ , } \;\;
4\pi \left(\frac{16 \pi}{3}\right)  \;\; \mbox{ and } \;\;
4\pi \left(\frac{16 \pi}{3}\right)^2  .                 \label{6-14}
\end{equation}

The sum (\ref{6-1}) over different Hilbert spaces $H^s_L$, equal to the
compound Hilbert space $H_S$, offers for each value of the 4-momentum a 
number of states with a multiplicity given by (\ref{6-14}).
A momentum eigenstate is therefore, in general, a superposition of states of 
the same 4-momentum, but belonging to different Hilbert spaces.
The momentum operator of the compound system is 
\begin{equation}
\sum_s p^s_\mu , \label{6-15}
\end{equation}
where $p^s_\mu$ is the momentum operator in $H^s_L$.
In each $H^s_L$, the operator $p^s_\mu$ satisfies a Dirac equation with the same 
mass $m_e$.
The operator (\ref{6-15}), therefore, satisfies a Dirac equation for the compound
system with a mass equal to $m_e$ multiplied by one of the multiplicities 
(\ref{6-14}). 
This reproduces the mass relations of G-M (\ref{6-8}).

The fact that the mass relations closely agree with the experimental data 
suggests that we have to identify the three configurations with the empirical 
massive leptons.

\section{Proton mass}

G-M obtained another mass from a decomposition of SO(3,3)
with respect to the product group SO(3,1)$\times$SO(2) \cite{ggm1}.
He calculated the volume factor
\begin{equation}
V\left(\frac{SO(3,3)}{SO(3,1) \times SO(2)}\right) = 2^5 \pi^6 , \label{7-1}
\end{equation}
or, relative to the volume factor of the electron (\ref{6-7}),  
\begin{equation}
6 \pi^5 = 1836.1185 . 
\end{equation}
This value is remarkably close to the empirical ratio of the proton 
and electron masses 
\begin{equation}
m_p / m_e = 1836.15267245(75) .
\end{equation}
The same result was obtained earlier by Wyler \cite{aw}.

The SO(3,3), obviously, originates from the same product group SO(3)$\times$SO(3) 
that has led us to the Poincar\'e group.
Now however, there is no restriction of the second SO(3) to the geometry of
the ecliptic.

Again, the close agreement with the experimental proton mass suggests that the
structure referred to by the factor (\ref{7-1}) corresponds to the empirical
proton.

\section{Baryonic structures}

In his already cited article \cite{ph}, O'Hara studied possible correlations 
of n (two-component) spinors:
``n particles are isotropically spin-correlated, if a measurement made in an 
ARBITRARY direction $\theta$ on ONE of the particles allows us to predict with 
certainty, the spin value of each other of the $n-1$ particles for the same
direction $\theta$''.

O'Hara then showed, that 
\begin{equation}
\left|\psi\right> = \frac{1}{\sqrt{2}}\left(\left|u\right>\left|u\right> 
+ \left|d\right>\left|d\right>\right)
\end{equation}
and
\begin{equation}
\left|\psi\right> = \frac{1}{\sqrt{2}}\left(\left|u\right>\left|d\right> 
- \left|d\right>\left|u\right>\right)
\end{equation}
are the only isotropically spin-correlated states permitted for a system
of n particles.

This means that when three spinors are coupled, only two of them may form 
a spin singlet or a spin triplet: the third spinor must be statistically 
independent.
In the case of singled coupling, the indistinguishability of the particles 
then forces the spinors into the state 
\begin{eqnarray}
\psi[\lambda_1, \lambda_2, \lambda_3] = \frac{1}{\sqrt{3}} 
\!\!\!\!\!&[&\!\!\!\!\psi_{12}[\lambda_1, \lambda_2]\,\psi_3(\lambda_3) \nonumber \\
&+&\!\!\!\!\psi_{31}[\lambda_1, \lambda_2]\,\psi_2(\lambda_3) \nonumber \\
&+&\!\!\!\!\psi_{23}[\lambda_1, \lambda_2]\,\psi_1(\lambda_3)\;] ,
\end{eqnarray}
where
\begin{equation}
\psi_{ij}[\lambda_1, \lambda_2] = \frac{1}{\sqrt{2}}
\left(\psi_i(\lambda_1) \psi_j(\lambda_2) - \psi_i(\lambda_2) \psi_j(\lambda_1) \right),
\end{equation}
which suggests that the spinors ``are in a dynamic equilibrium with each 
other, with the coupling continuously broken and then reformed among 
different (spinors).''

The ``dynamic equilibrium'' may be phenomenologically explained as the result 
of an interaction.
In contrast to the electromagnetic interaction, which in \cite{ws} has been
explained as the result of the irreducibility of two-particle states, this 
interaction results from an extension of Pauli's principle to a three-spinor 
configuration.

O'Hara concluded: ``if it is assumed that only singlet state coupling is stable, 
then all spin-3/2 configurations will necessarily decompose,'' whereas spin-1/2
states will be stable.
This closely resembles the structures of spin-1/2 and spin-3/2 baryons.
``However, [now] their ... structure may be explained in terms of the 
coupling principle, without any recourse to the concept of color.''

\section{Conclusions}

Sets of binary elements exhibit an unexpectedly rich internal structure, emerging 
from two basic symmetries:\\
$\bullet$ Symmetry with respect to interchanging the states ``up'' and ``down'' of a binary 
element (isotropy), and\\
$\bullet$ Symmetry with respect to interchanging two binary elements (indistinguishability).\\
The first symmetry has been used to establish a Poincar\'e invariant parameter space.
The independence from any specific frame of reference then results in a quantum mechanical 
description.
In this description, single binary elements possess all the characteristics of Dirac spinors,
including a mass spectrum close to the empirical leptonic mass spectrum.
The second symmetry leads to the Pauli exclusion principle and to baryonic structures. 

In a previous article \cite{ws}, the author has shown that two Dirac spinors exhibit 
an interaction, with the properties of the electromagnetic interaction, when they are 
described by an irreducible two-particle representation of the Poincar\'e group.
In another article \cite{ws1}, the author has demonstrated how space-time and
gravity emerge from the basic properties of Poincar\'e invariant quantum 
mechanics.

All these results have been obtained by mathematical deductions from the 
basic symmetries of binary elements---without any reference to ``laws 
of nature'' or to ``natural constants.''

This raises the question about the meaning of ``physical reality.'' 
When the basic structures of theoretical particle physics can be derived solely 
from properties inherent in the concept of binary information, then what we comprehend 
as physical reality is nothing other than a reflection of some predetermined 
informational structures that we take advantage of in order to collect and categorize 
information about reality.
Then the physics of elementary particles is basically a ``physics of binary 
information.''

Einstein's quote, ``It is the theory which decides what can be observed'' 
\cite{aeh} perfectly illustrates the role of the physics of binary information:
It tells us what we see when we look at reality on the most elementary 
level of information, formed by binary information.
Beyond this level, there is no further information.
The physics of binary information, therefore, marks the basis of physics.

\end{document}